\documentclass[twocolumn,showpacs,superscriptaddress,amsmath,amssymb,prc,preprintnumbers]{revtex4-1}
\usepackage[pdftex]{graphicx}% Include figure files
\usepackage{dcolumn}% Align table columns on decimal point
\usepackage{bm}% bold math
\usepackage{ifpdf}
\usepackage{epstopdf}
\usepackage{slashed}
\usepackage{amsfonts}
\usepackage{mathrsfs}
\usepackage{amssymb}
\usepackage{multirow,longtable}
\usepackage{CJK}
\usepackage[pdftex]{xcolor}
\allowdisplaybreaks[4]
\begin{document}
\begin{CJK*}{UTF8}{}

\title{The description of giant dipole resonance key parameters with multitask neural networks}

\author{J. H. Bai (\CJKfamily{gbsn}{白景虎})}
\affiliation{School of Nuclear Science and Technology, Lanzhou University, Lanzhou 730000, China}

\author{Z. M. Niu (\CJKfamily{gbsn}{牛中明})}
\affiliation{School of Physics and Materials Science, Anhui University, Hefei 230601, China}
\affiliation{Institute of Physical Science and Information Technology, Anhui University, Hefei 230601, China}

\author{B. Y. Sun (\CJKfamily{gbsn}{孙保元})}
\affiliation{School of Nuclear Science and Technology, Lanzhou University, Lanzhou 730000, China}

\author{Y. F. Niu (\CJKfamily{gbsn}{牛一斐})}
\email{niuyf@lzu.edu.cn}
\affiliation{School of Nuclear Science and Technology, Lanzhou University, Lanzhou 730000, China}

\begin{abstract}
Giant dipole resonance (GDR) is one of the fundamental collective excitation modes in nucleus. Continuous efforts have been made to the evaluation of GDR key parameters in different nuclear data libraries.
We introduced multitask learning (MTL) approach to learn and reproduce the evaluated experimental data of GDR key parameters,  including both GDR energies and widths. Compared to the theoretical GDR parameters in RIPL-3 library, the accuracies of MTL approach are almost doubled for 129 nuclei with experimental data. The significant improvement is largely due to the right classification of unimodal nuclei and bimodal nuclei by the classification neural network.  Based on the good performance of the neural network approach, an extrapolation to 79 nuclei around the $\beta$-stability line without experimental data is made, which provides an important reference to future experiments and data evaluations. The successful application of MTL approach in this work further proofs the feasibility of studying multi-output physical problems with multitask neural network in nuclear physics domain.
\end{abstract}
%
%
%%\begin{keyword}
%%  Giant dipole resonance \sep peak energy \sep resonance width  \sep multitask neural network
%%\end{keyword}
\maketitle
\end{CJK*}

\section{Introduction}

Giant dipole resonance (GDR) is one of the fundamental modes of nuclear collective excitations, whose energy exceeds the binding energy of nucleons \cite{Plujko2018,Kleinig2008}.
The study of GDRs can contribute to the understanding of nuclear structure, for example, the ground-state deformation of nuclei can be reflected by the shape of the GDR strength distributions \cite{Maruhn2005, Myers1977, Goriely2019}.
Furthermore, GDR is a collective dipole oscillation of protons versus neutrons giving rise to a dynamic electric-dipole (E1) moment \cite{Savran2006}, which reflects asymmetry information in nuclear equation of state (EoS) \cite{Bracco2019}.

The experimental GDR data have been measured in various types of experiments, namely, photonuclear experiments \cite{Kawano2020109} with photons from bremsstrahlung radiation \cite{Martinez1995}, positron annihilation in flight \cite{Jones1999}, and more recently laser Compton scattering (LCS) \cite{Tanaka2020,GuoWei2008}, as well as ($p,p'$) reaction \cite{Tamii2011}, and so on.
The experimental GDR data can be fitted using the Lorentzian curve, from which the key GDR parameters that  consist of the resonance energy $E$ and shape width $\Gamma$ can be derived.
In major photonuclear data libraries (RIPL \cite{Capote2009}, IAEA \cite{Kawano2020109}, CENDL \cite{cendl} etc.), experimental data of most nuclei near the $\beta$-stability line are available.
In recent years, a new international coordinated research project (CRP) has been launched by IAEA, reevaluating the GDR experimental data, in order to improve the reliability of experimental data and to address the growing needs for photonuclear data \cite{Plujko2018,Kawano2020109, Goriely2019}.
Up to now, according to the International Atomic Energy Agency (IAEA) photonuclear data library, 219 isotopes were evaluated, including revisited 164 isotopes in the previous library \cite{Plujko2018,Kawano2020109}, 37 isotopes with newly available experimental data, and 18 isotopes evaluated by a model prediction.

The increase of GDR experimental data brings challenges to relevant theoretical models.
Nowadays two types of nuclear models are mainly used in GDR predictions: microscopic approaches and phenomenological approaches.
In microscopic approaches, quasiparticle random phase approximation (QRPA) method is used most frequently.
In recent years, QRPA models for the study of GDRs  have been implemented fully self-consistently based on various density functionals, such as Skyrme functional \cite{Goriely2002,Goriely2004}, Gogny functional \cite{Peru200744,Goriely2018,Martini2011} and relativistic functional \cite{Paar2003,Paar2007}. As a microscopic model, QRPA achieved great success in describing the centroid energies of GDRs, however, it fails to describe the resonance width. To overcome this problem, beyond RPA approaches, such as second RPA \cite{Gambacurta2010,Grasso2020} and RPA with particle vibration coupling effects \cite{Litvinova2007,Egorova2016,Roca-Maza2017}, were developed. However, due to the big computation cost, these models still haven't been used for large-scale calculations.
On the other hand, in the phenomenological approaches, the photoabsorption cross-section of GDR is usually described by the Lorentzian representation, such as the standard Lorentzian (SLO) model \cite{DMBrink195501, Axel1962}, the modified Lorentzian (MLO) approach \cite{Plujko2002123}, and its simplified version SMLO \cite{Plujko20071720}, and so on.
Besides, based on the phenomenological Goldhaber and Teller (GT) model \cite{Goldhaber1948,Goriely199810}, there are predictions of the GDR energies and widths for about 6000 nuclei with $14\leq Z \leq110$ lying within the proton and the neutron driplines, which are compiled in RIPL-3 as theoretical GDR parameters \cite{Capote2009,RIPL3}.
Those microscopic and phenomenological models have given a reliable overall description of GDR, especially in medium to heavy mass nuclei.
However, there is still a large space for improvement between theoretical results and  experimental data.

With the development of computation techniques, machine learning shows its great power in learning complex big data and making predictions, and also began to show its usefulness in nuclear domain in recent years.
Many fundamental properties of atomic nuclei have been explored by using machine learning methods.
Bayesian neural network (BNN) was successfully used for the accurate descriptions of  nuclear masses \cite{Niu201848,Niu20198,Utama2016,Neufcourt2018}, fission yields of actinide nuclei \cite{WangZiAo2019}, as well as $\beta$-decay half-lives \cite{NiuZM2019}.
The nuclear mass predictions are also studied by introducing the Fourier spectral analysis \cite{Niu2018scibull} and the radial basis function (RBF) approach \cite{Niu2016}.
The $\beta$-decay half-lives are also studied by a fully connected, multilayer artificial neural network \cite{Costiris2009,Costiris2013}.
These previous studies inspired us to use  machine learning for the study of GDR key parameters, including resonance energies and widths. However, being different with previous applications in nuclear physics, this problem has two typical characters: (i) it is a multi-output problem, with both energies and widths as outputs; (ii) the data amount of GDR parameters is relatively small, where we select 129 isotopes with experimental data of GDR parameters.  Regarding these characters, multitask learning (MTL) \cite{Caruana001, zhang2017survey} approach, which has not been applied in nuclear physics,  is an ideal tool for the present problem. MTL approach can resolve multi-output problems very well by introducing multiple loss functions, which can get the optimal solution for each output, and avoid the situations where one output is overfitting and the other outputs have not yet reached the best-fitting state. Thus, MTL can be a good solution to deal with the problem which has multiple related tasks each of which has limited learning data \cite{zhang2017survey}.

In this work, we will introduce a new machine learning approach for the study of  GDR key parameters, including resonance energies and widths.
We combine three networks to build the model for GDR, which consists of one traditional classification neural network and two MTL networks. The GDR data are divided into two groups according to the number of peaks (single or double) by the classification neural network, and two MTL networks are then used for data training of each group. The machine learning results will be compared with those from GT model in RIPL-3 library \cite{RIPL3} to see how much improvement has been made. Particular attention is paid to the correct predictions of single or double peaks of GDR.

\section{The Models}

The experimental photo absorption cross section of GDR can be well fitted by Lorentzian functions, from which a group of GDR parameters can be obtained for each nucleus. SLO \cite{DMBrink195501, Axel1962} model is one of the most frequently used methods to obtain the experimental GDR parameters in the evaluation of photonuclear data.

Following Ref. \cite{Plujko2018}, for a photon with energy $\varepsilon_{\gamma}$, the photoabsorption cross section $\sigma_{\rm abs}(\varepsilon_{\gamma}) $ is taken as a sum of the terms corresponding to the GDR excitation given by $\sigma_{\rm GDR}(\varepsilon_{\gamma})$ and the quasi-deuteron photodisintegration $\sigma_{\rm QD}(\varepsilon_{\gamma})$ \cite{Chadwick2000, Chadwick1991},
\begin{align}
\sigma_{\rm abs}(\varepsilon_{\gamma}) = \sigma_{\rm GDR}(\varepsilon_{\gamma}) + \sigma_{\rm QD}(\varepsilon_{\gamma}).
\end{align}

The expression for the minor contribution $\sigma_{\rm QD}(\varepsilon_{\gamma})$ can be found in Ref. \cite{Plujko2018}, and here we mainly discuss the dominated component $\sigma_{\rm GDR}(\varepsilon_{\gamma})$.
$\sigma_{\rm GDR}(\varepsilon_{\gamma})$ is fitted by the SLO model.
\begin{eqnarray}
\sigma_{\rm GDR}(\varepsilon_{\gamma}) & = & \sum_{j=1}^{j_m} \sigma_{{\rm GDR}, j}(\varepsilon_{\gamma})  = \sigma_{\rm TRK} s_j \cdot F_j(\varepsilon_{\gamma}),\\
F_j(\varepsilon_{\gamma}) &= & \frac {2}{\pi} \frac{\varepsilon_{\gamma}^2\Gamma_j}{[(\varepsilon_{\gamma}^2-(E_{j})^2]^2+[\varepsilon_{\gamma}\Gamma_j]^2} , \\
\sigma_{\rm TRK} &= & 60 \frac {NZ}{A}.
\end{eqnarray}

The GDR parameters in SLO model consist of the resonance energy $E_{j}$ and shape width $\Gamma_{j}$ of the $j$-th mode of the giant dipole excitation
for one-component Lorentzian nuclei with $j_m = 1$ and two-component Lorentzian nuclei with $j_m = 2$. Usually, deformed nuclei are calculated by two-component Lorentzian, since for deformed nuclei, the protons and neutrons oscillate against each other parallel to the axis of rotational symmetry as well as perpendicular to it.
$s_j $ is the normalized contribution of the Lorentzian component $F_j $ in terms of the Thomas-Reiche-Kuhn (TRK) sum rule $\sigma_{\rm TRK}$.
 $\Gamma_j  $ is the GDR width, which is a constant that doesn't depend on the $\gamma$-ray energy.

To obtain the GDR parameters $E_j$ and $\Gamma_j$ by machine learning, we built three neural  networks. As the first step, we should distinguish the one-component Lorentzian nuclei with $j_m = 1$ and two-component Lorentzian nuclei with $j_m = 2$, which was accomplished by the classification neural network (Net$_0$). Considering the effect of deformation parameter $\beta_2$ on the shape of Lorentzian curve, we will take $\beta_2$ as one of the inputs in the input layer, together with proton number $Z$, neutron number $N$ and mass number $A$ of the nucleus, i.e., $\boldsymbol{x} = (Z, N, A, \beta_2)$. The output of Net$_0$ is  $j_m$, which is 1 or 2 representing the number of Lorentzian components.
The inputs are connected with the outputs through the Net$_0$:
\begin{align}
&\boldsymbol{a}^{(1)} = {\rm tanh}(\boldsymbol{\theta}^{(1)} \boldsymbol{x}  + \boldsymbol{b}^{(1)}) ,
\boldsymbol{a}^{(2)} = {\rm tanh}(\boldsymbol{\theta}^{(2)} \boldsymbol{a}^{(1)} +\boldsymbol{b}^{(2)}) \notag\\
&\boldsymbol{S}_{\boldsymbol{\theta},\boldsymbol{b}} (\boldsymbol{x}) = \boldsymbol{a}^{(3)} = {\rm tanh}(\boldsymbol{\theta}^{(3)} \boldsymbol{a}^{(2)} + \boldsymbol{b}^{(3)})
\end{align}
where tanh is the activation function, which provides the nonlinearity for the net.
The $\boldsymbol{a}^{(1)}$ and $\boldsymbol{a}^{(2)}$ are the first and second hidden layer.
The layer $\boldsymbol{a}^{(3)}$ is the hypothetical output $\boldsymbol{S}_{\boldsymbol{\theta},\boldsymbol{b}} $ of the Net$_0$. The dimensions $L$ of the vectors $\boldsymbol{x}, \boldsymbol{a}^{(i)}$  are
\begin{align}
L(\boldsymbol{x}) = 4 , L(\boldsymbol{a}^{(1)}) =L(\boldsymbol{a}^{(2)})= 6 , L(\boldsymbol{a}^{(3)}) = 1.
\end{align}
The dimension of the weight matrices $\boldsymbol{\theta}^{(i)}$ is $L(\boldsymbol{a}^{(i)}) \times L(\boldsymbol{a}^{(i-1)})$ if we label $\boldsymbol{a}^{(0)} = \boldsymbol{x}$, and the dimension of the bias parameters $L(\boldsymbol{b}^{(i)})= L(\boldsymbol{a}^{(i)})$, where $i=1,2,3$.  Since $L(\boldsymbol{S}_{\boldsymbol{\theta},\boldsymbol{b}})=1$, we will use $S_{\boldsymbol{\theta},\boldsymbol{b}} (\boldsymbol{x})$ instead.

The loss function $J(\boldsymbol{\theta},\boldsymbol{b})$, which is used to describe the difference between the hypothetical output $S_{\boldsymbol{\theta},\boldsymbol{b}} (\boldsymbol{x})$ and  the number of Lorentzian components $j_{mi}$, can be obtained through
\begin{align}
J(\boldsymbol{\theta},\boldsymbol{b}) &= \frac{1}{2N}\sum_{i=1}^N (S_{\boldsymbol{\theta},\boldsymbol{b}} (\boldsymbol{x}_i)-j_{mi})^2,
\end{align}
where $N$ is the number of data.
\begin{align}
\boldsymbol{\theta}_{0},\boldsymbol{b}_{0} &= \mathop{minimize}\limits_{\boldsymbol{\theta},\boldsymbol{b}}  J(\boldsymbol{\theta},\boldsymbol{b}).
\end{align}

The optimal parameter set $\boldsymbol{\theta}_{0},\boldsymbol{b}_{0}$ of Net$_0$ can be obtained by using the Adam optimizer \cite{kingma2014adam} to minimize the loss function. As a result, the accuracy of classification of one-component Lorentzian (unimodal) nuclei and two-component Lorentzian (bimodal) nuclei  reaches about 98\%. After the classification, we shall use MTL networks Net$_1$ and Net$_2$ for obtaining GDR parameters for each group of nuclei.

Net$_1$ is used to train the GDR parameters of unimodal nuclei, i.e., $E$ and $\Gamma$, while Net$_2$ is used to train the GDR parameters of bimodal nuclei, i.e., ($E_1, E_2$) and ($\Gamma_1, \Gamma_2$). The inputs for Net$_1$ and Net$_2$ are the same as those for Net$_0$, which are $\boldsymbol{x} = (Z, N, A, \beta_2)$.
The structure of Net$_1$ is as follows,
\begin{align}
&\boldsymbol{a}^{(1)} = {\rm tanh}(\boldsymbol{\theta}^{(1)} \boldsymbol{x}  + \boldsymbol{b}^{(1)}) ,\notag\\
&\boldsymbol{a}^{(2)} = {\rm tanh}(\boldsymbol{\theta}^{(2)} \boldsymbol{a}^{(1)}  + \boldsymbol{b}^{(2)}) ,
\boldsymbol{a}^{(3)} = {\rm tanh}(\boldsymbol{\theta}^{(3)} \boldsymbol{a}^{(1)} + \boldsymbol{b}^{(3)}) ,\notag\\
&\boldsymbol{a}^{(4)} = {\rm tanh}(\boldsymbol{\theta}^{(4)} \boldsymbol{a}^{(2)} + \boldsymbol{b}^{(4)}) ,
\boldsymbol{a}^{(5)} = {\rm tanh}(\boldsymbol{\theta}^{(5)} \boldsymbol{a}^{(3)} + \boldsymbol{b}^{(5)}) ,\notag\\
&\boldsymbol{S}^E_{\boldsymbol{\theta},\boldsymbol{b}}(\boldsymbol{x}) = \boldsymbol{a}^{(4)}, \boldsymbol{S}^\Gamma_{\boldsymbol{\theta},\boldsymbol{b}}(\boldsymbol{x}) = \boldsymbol{a}^{(5)} ,
\end{align}
with dimensions of the above vectors being
\begin{align}
&L(\boldsymbol{a}^{(1)}) =L(\boldsymbol{a}^{(2)})=L(\boldsymbol{a}^{(3)})= 4 ,L(\boldsymbol{a}^{(4)}) =L(\boldsymbol{a}^{(5)})=1.
\end{align}

The layers $\boldsymbol{a}^{(2)}$ and $\boldsymbol{a}^{(3)}$ are both calculated from $\boldsymbol{a}^{(1)}$, but separated to calculate each output. Thus  $\boldsymbol{a}^{(1)}$ is called  shared layer, while $\boldsymbol{a}^{(2)}$ and $\boldsymbol{a}^{(3)}$ are called task layers.
The outputs of Net$_1$ are $\boldsymbol{a}^{(4)}$ and $\boldsymbol{a}^{(5)}$.
Their loss functions are defined as follows,
\begin{align}
J^E(\boldsymbol{\theta},\boldsymbol{b}) &= \frac{1}{2N}\sum_{i=1}^N \frac{ (S^E_{\boldsymbol{\theta},\boldsymbol{b}}(\boldsymbol{x}_i)-y^E_{i})^2 }{(t^E_{i} + \bar{t}^E)^2}, \label{align:align11}
\\
J^{\Gamma}(\boldsymbol{\theta},\boldsymbol{b}) &= \frac{1}{2N}\sum_{i=1}^N \frac{ (S^{\Gamma}_{\boldsymbol{\theta},\boldsymbol{b}}(\boldsymbol{x}_i)-y^\Gamma _{i})^2 }{( t^\Gamma_{ i} + \bar{t}^\Gamma)^2}, \label{align:align12}
\end{align}
where $N$ is the number of data; $y^E_{i}$ and $y^\Gamma_{ i}$ are experimental data of energy $E$ and width $\Gamma$; $t^E_{i}$ and $t^\Gamma_{i}$ are the corresponding experimental errors.
The introduction of the averaged experimental errors for energy and width $\bar{t}^E$ and $\bar{t}^\Gamma$  is to avoid the divergence caused by small experimental errors that close to zero, which are defined as $\bar{t}^E = \frac{1}{N} \sum_{i=1}^{N}t^E_i $, $\bar{t}^\Gamma = \frac{1}{N} \sum_{i=1}^{N}t^\Gamma_i $.

\begin{figure}[htbp]\setlength{\abovecaptionskip}{0.0em}
\includegraphics[width = 0.99\linewidth]{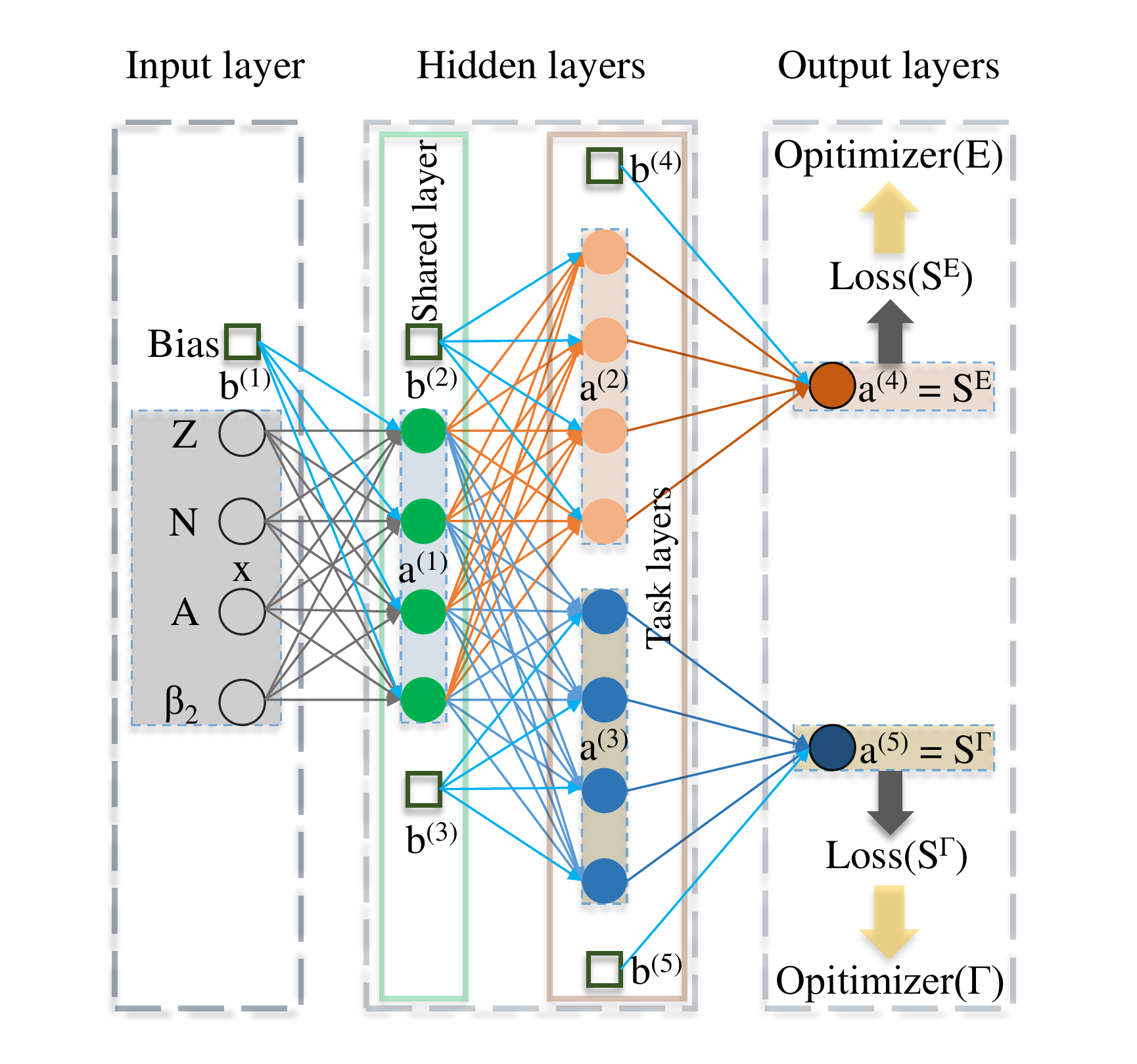}
\caption{A schematic diagram of the structure of MTL network Net$_1$ for computing unimodal nuclei.}\label{fig:fig1}
\end{figure}

The structure of Net$_1$ is shown in Fig. \ref{fig:fig1}.
It's clear to see that each output has its own task layer, loss function, as well as optimizer, so as a result each output is calculated optimally.
Moreover, the shared layer implies the relationship between tasks, making communications among different tasks possible.

The structure of Net$_2$ is similar to that of Net$_1$, however, Net$_2$ has 4 outputs so that there are 4 task layers and 2 shared layers, making the network more complicated.

The structure of Net$_2$ is shown as follows,
\begin{align}
&\boldsymbol{a}^{(1)} = {\rm tanh}(\boldsymbol{\theta}^{(1)} \boldsymbol{x} + \boldsymbol{b}^{(1)}) , \boldsymbol{a}^{(2)} = {\rm tanh}(\boldsymbol{\theta}^{(2)} \boldsymbol{a}^{(1)} + \boldsymbol{b}^{(2)}), \notag\\
&\boldsymbol{a}^{(3)} = {\rm tanh}(\boldsymbol{\theta}^{(3)} \boldsymbol{a}^{(2)} + \boldsymbol{b}^{(3)}) , \boldsymbol{a}^{(4)} = {\rm tanh}(\boldsymbol{\theta}^{(4)} \boldsymbol{a}^{(2)} + \boldsymbol{b}^{(4)}), \notag\\
&\boldsymbol{a}^{(5)} = {\rm tanh}(\boldsymbol{\theta}^{(5)} \boldsymbol{a}^{(2)} + \boldsymbol{b}^{(5)}) , \boldsymbol{a}^{(6)} = {\rm tanh}(\boldsymbol{\theta}^{(6)} \boldsymbol{a}^{(2)} + \boldsymbol{b}^{(6)}), \notag\\
&\boldsymbol{a}^{(7)} = {\rm tanh}(\boldsymbol{\theta}^{(7)} \boldsymbol{a}^{(3)} + \boldsymbol{b}^{(7)}) ,\boldsymbol{a}^{(8)} = {\rm tanh}(\boldsymbol{\theta}^{(8)} \boldsymbol{a}^{(4)} + \boldsymbol{b}^{(8)}), \notag\\
&\boldsymbol{a}^{(9)} = {\rm tanh}(\boldsymbol{\theta}^{(9)} \boldsymbol{a}^{(5)} + \boldsymbol{b}^{(9)}) ,\boldsymbol{a}^{(10)} = {\rm tanh}(\boldsymbol{\theta}^{(10)} \boldsymbol{a}^{(6)} + \boldsymbol{b}^{(10)}), \notag\\
&\boldsymbol{S}^{E_1} = \boldsymbol{a}^{(7)} ,   \boldsymbol{S}^{E_2} = \boldsymbol{a}^{(8)} , \boldsymbol{S}^{\Gamma_1} = \boldsymbol{a}^{(9)} ,  \boldsymbol{S}^{\Gamma_2} = \boldsymbol{a}^{(10)},
\end{align}
with dimensions of above vectors being
\begin{align}
&L(\boldsymbol{a}^{(1)}) =L(\boldsymbol{a}^{(2)})=L(\boldsymbol{a}^{(3)})= \cdots =L(\boldsymbol{a}^{(6)})= 4, \notag\\
&L(\boldsymbol{a}^{(7)}) =L(\boldsymbol{a}^{(8)}) =L(\boldsymbol{a}^{(9)}) =L(\boldsymbol{a}^{(10)}) =1.
\end{align}
Net$_1$ and Net$_2$ use the Adam optimizer to minimize the loss function.

The root-mean-square (rms) deviations from experimental data of calculated GDR peak energies  $\sigma_{\rm rms}(E)$ are obtained by
\begin{align}
\sigma_{\rm rms}(E) =   \sqrt{ \frac{1}{2N} \sum_{i=1}^N  [ (E_{1i}^{ \rm cal}- y^{E_1}_{i})^2  +  (E_{2i}^{ \rm cal} - y^{E_2}_{i})^2 ] }. \label{align:align13}
\end{align}
$E_{1i}^{ \rm cal}$ and $E_{2i}^{ \rm cal}$ are calculated energies of first peak and second peak respectively. In the case of neural network, $E_{1i}^{ \rm cal} = S^{E_1}_{\boldsymbol{\theta},\boldsymbol{b}}(\boldsymbol{x}_i)$, and  $E_{2i}^{ \rm cal} = S^{E_2}_{\boldsymbol{\theta},\boldsymbol{b}}(\boldsymbol{x}_i)$.
For unimodal nuclei, we just take $E_{2i}^{ \rm cal} =  y^{E_2}_{i}= 0$ in the calculation of $\sigma_{\rm rms}(E)$. For the case of unimodal nuclei misjudged as bimodal nuclei in theory, we only consider the first peak from theory in the comparison with experimental data; for the opposite case, we use $E_{1i}^{ \rm cal}=E_{2i}^{ \rm cal}=E_{i}^{ \rm cal}$.   The rms deviations of GDR resonance widths by neural network $\sigma_{\rm rms}(\Gamma)$  are calculated in the same way as $\sigma_{\rm rms}(E)$.

The deformation parameters $\beta_2$ in the input are a combination of experimental data \cite{RIPL2} when available and finite-range liquid-drop model (FRDM) results \cite{Moller2016}.
The experimental GDR parameters to be trained are taken from the results fitted by SLO model in International Atomic Energy Agency Photonuclear Data Library 2019
(IAEA2019) \cite{Kawano2020109,Plujko2018}. The nuclei with experimental errors $\sigma ^{\rm exp} < 1.5$ MeV  for the widths are considered.
As a result there are 129 nuclei left and hence 366 data including GDR energies and widths, which compose the entire data set.
In order to examine the validity of the MTL approach, we separate the entire data set into the learning set and the validation set with a ratio of about 9:1.
The learning set is built by randomly selecting 116 nuclei from the entire set, and the remaining 13 nuclei compose the validation set.
We have tested the sensitivity of the trained MTL model with respect to hyper-parameters, including learning rate and sizes of training and validation data-sets. By changing the learning rate from the optimal value 0.008 to 0.016 and 0.001, the change of rms deviations of GDR energies and widths from experimental data are generally within 9\%. In addition, if we increase the data-set ratio between validation and learning data-sets from 1:9 to 3:7, the change of rms deviations is still within 18\%. Therefore, the results are not very sensitive to the hyper-parameters, which further implies the reliability of the present networks to predict the GDR energies and widths.

In the results and discussions, we will compare the results from MTL approach with those from the GT model. The GT model is referring to the theoretical results used in RIPL-3 library \cite{RIPL3}. It is calculated  based on the Goldhaber-Teller model \cite{Goldhaber1948} where the neutron and proton densities perform an out-of-phase vibration around their center of mass. The dynamics of the oscillation is assumed to be dominated by the np-interaction \cite{Isacker1992}. The strength of np-interaction is derived from a least-square fit to the experimental GDR energies \cite{Goriely199810}. The nucleon density distribution and ground-state deformation are taken from the Extended Thomas-Fermi plus Strutinsky Integral (ETFSI) compilation \cite{Aboussir1995}. The expression for the shell-dependent GDR width is taken from \cite{Thielemann1983} using the newly-determined GDR energies and the ETFSI shell corrections. Compared to the network approach, the GT model has much less free parameters due to a clear physical picture. Being a completely different approach, the neural network is featured by a large number of connected parameters that extract the modularized information from data. For the MTL networks, the numbers of parameters are 70 for Net$_1$ and 140 for Net$_2 $. One way to assess if the parameter set is too big is to check if the network is overfitted. We have checked that our neural networks are not overfitted by comparing the rms deviations between learning set and validation set, which give similar values for both data-sets. In addition, from the rms deviations of our results with respective to the data in learning set and validation set, it is also found that the present networks give the best results compared to other networks with fewer or more parameters.

\section{Results and discussions}

\begin{figure}[htbp]\setlength{\abovecaptionskip}{0.0em}
\includegraphics[width = 0.99\linewidth]{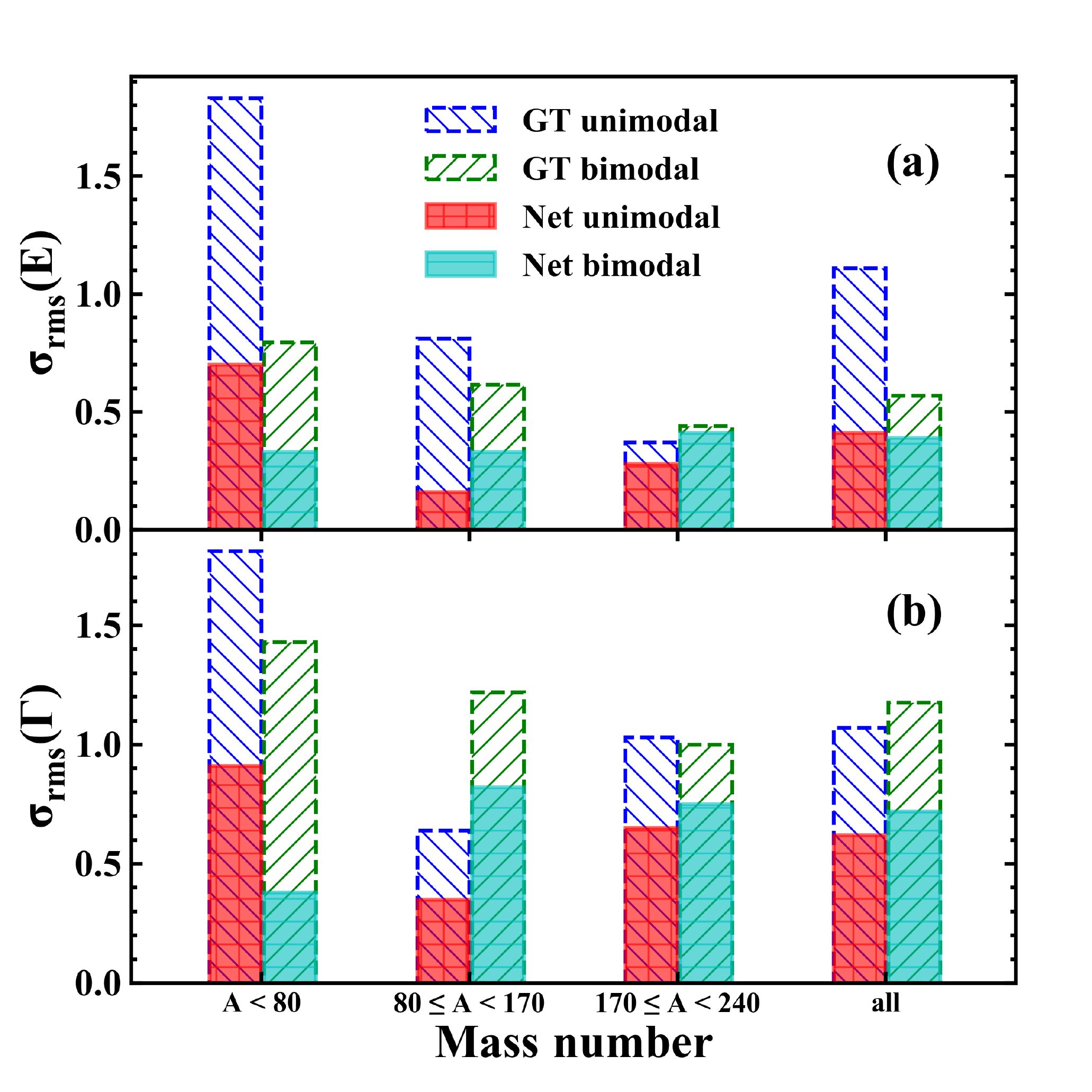}
\caption{ The root-mean-square (rms) deviations of GDR (a) peak energies and (b) resonance widths  with respect to experimental data from the IAEA library \cite{Plujko2018} for results from GT model \cite{RIPL3} and MTL networks. The data set is divided into three mass regions, and
  further classified by one-component Lorentzian (unimodal) nuclei and two-component Lorentzian (bimodal) nuclei in each mass region.}\label{fig:fig2}
\end{figure}

\begin{figure}[htbp]\setlength{\abovecaptionskip}{0.0em}
\includegraphics[width = 0.99\linewidth]{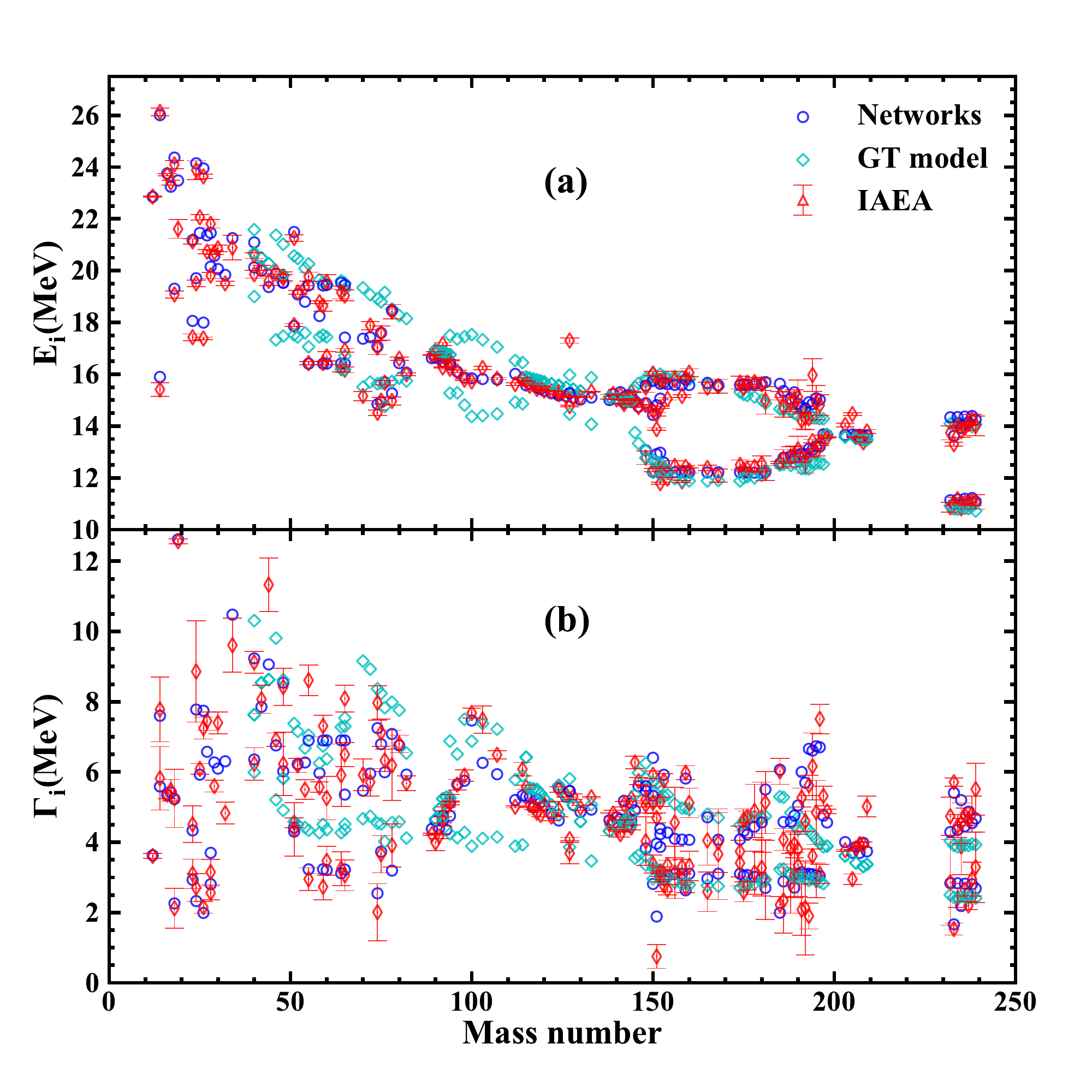}
\caption{The key parameters of GDR, (a) peak energies and (b) resonance widths, as functions of nuclear mass number $A$, calculated by MTL networks (blue circle) and GT model \cite{RIPL3} (cyan square), in comparison with experimental data from the IAEA library \cite{Plujko2018} (red diamond). }\label{fig:fig3}
\end{figure}

In Fig. \ref{fig:fig2}, the root-mean-square (rms) deviations of GDR  peak energies and  resonance widths  with respect to experimental data for results from GT model and MTL networks are compared. To analyze these results in details, the comparisons  are done for unimodal nuclei and bimodal nuclei in three mass regions respectively. Generally, it is clearly seen that the MTL networks improved both GDR  peak energies and  resonance widths significantly compared to GT model. For the peak energy, the rms deviation is reduced by 51.2\% from GT model to MTL approach for all nuclei, while this number is 41.4\% for resonance width. To understand the different levels of improvements for energies and widths, we further check the results of Net$_1$ and Net$_2$ respectively.  For unimodal nuclei, Net$_1$ improves the rms deviation by 63.1\% for energies and 42.1\% for widths,  while for bimodal nuclei, Net$_2$ improves the rms deviation by 31.6\% for energies and 38.7\% for widths, seen in Fig. \ref{fig:fig2}. The improvement for energy in Net$_1$ is much bigger than that for widths, in fact due to that the GT model gives a very poor description of energies for unimodal nuclei, which will be discussed in more details in panel (a) of Fig. \ref{fig:fig2} and Fig. \ref{fig:fig3}.

\begin{figure}[htbp]\setlength{\abovecaptionskip}{0.0em}
\includegraphics[width = 0.99\linewidth]{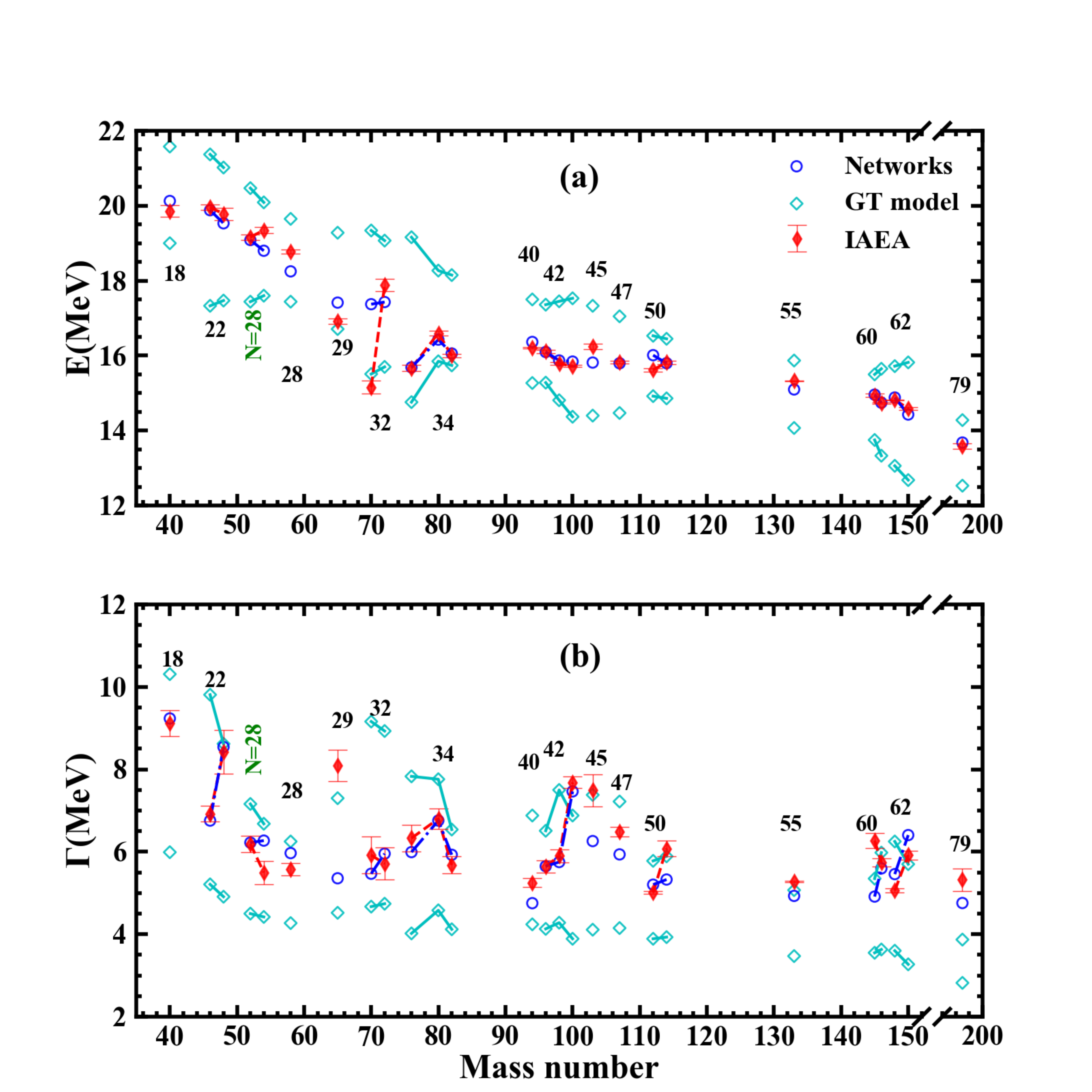}
\caption{The key parameters of GDR (a) peak energies and (b) resonance widths of the unimodal nuclei misjudged by GT model,  calculated by MTL networks (blue circle) and GT model \cite{RIPL3} (cyan square), in comparison with experimental data from the IAEA library \cite{Plujko2018} (red diamond). The same isotopes are linked by lines with the corresponding proton number shown aside, and the same isotones of $N=28$ are also linked. }\label{fig:fig4}
\end{figure}

Compared among different mass regions, the most significant improvement happens in the intermediate mass region $80\leq A<170$, due to the right classification of unimodal and bimodal nuclei in MTL approach, which will be discussed in details in Fig. \ref{fig:fig3} and Fig. \ref{fig:fig4}. The second big improvement is in the light mass region $A<80$, where the GT model gives the worst results compared to other mass regions. Based on the picture that protons and neutrons oscillate with each other, this classical model does not perform well for light nuclei. The MTL networks can overcome this problem to large extent, and give similar accuracies for the description of $A<80$ nuclei and heavier nuclei, except for the unimodal nuclei in light mass region. The relative poor description of unimodal nuclei with $A<80$ is still due to the fact that the collectivity of light nuclei is not as strong as that in heavier nuclei and thus shell effects tend to play their roles, leading to less good systematics in light nuclei, seeing Fig. \ref{fig:fig3} and Fig. \ref{fig:fig4}.

In panel (a) of Fig. \ref{fig:fig2}, it is noticed that improvements by MTL approach are much larger for unimodal nuclei than that for bimodal nuclei in all mass regions. In other words, it is apparent that in the GT model, the description for bimodal nuclei is better than that for unimodal nuclei, while in MTL approach, the descriptions of unimodal and bimodal nuclei reach similar accuracy except for light mass region with $A<80$. This interesting phenomenon will be explained in Fig. \ref{fig:fig3}.

The rms derivation of energies given by MTL approach is lower than 0.41 MeV, excluding the 24 unimodal nuclei with $A<80$, which shows a great performance of MTL approach. For unimodal nuclei with $A<80$, although the rms deviation of energies is higher, which is 0.70 MeV, it still obtains big improvement compared to the GT model. For the widths, the rms deviation is larger than that for energies in general, both for GT model and MTL approach, since widths have more complicated physical origins and worse systematics compared to energies. For MTL approach, the larger error bar of experimental data for GDR widths (seeing Fig. \ref{fig:fig3})  leads to smaller weights through Eq. \ref{align:align11} and \ref{align:align12} in the training process, and as a result, it further causes the worse description for GDR widths than for energies.

\begin{table}[h]
\caption{ The accuracies of classification of unimodal nuclei and bimodal nuclei for neural network approach and GT model in different mass regions.} \label{tab:accuracy}
\begin{tabular}{ccccc}
\hline \hline
      &$A<80$ &$80\leq A<170$&$170\leq A<240$& all  \\
\hline
GT &  60.0\% & 72.7\% & 97.1\% & 77.2\%     \\
Net$_0$  & 97.4\% & 98.2\% & 100.0\% &  98.5\%     \\
\hline \hline
\end{tabular}
\end{table}

In Fig. \ref{fig:fig3}, GDR energies and widths as functions of nuclear mass number calculated by MTL networks and GT model are shown, in comparison with experimental data from the IAEA library \cite{Plujko2018}.  It is clear  that  generally GDR peak energies ($E$) have good systematics with nuclear mass number $A$, which are approximately proportional to $A^{-1/3}$, as told in textbook.  However, for resonance widths, there is no clear evolution trend with mass number. It is also apparent that the experimental errors of resonance widths are much larger than that for peak energies. In the process of training networks, we take into account the experimental errors as the weights of data, which leads to smaller weights of resonance widths than that of peak energies. Together with the bad systematics, the learning of resonance widths is not as good as that of peak energies.

Comparing between two calculation results, it is clear that neural networks give better results than GT model does, especially for mass region with $A<150$. The large discrepancy between GT model and experimental data mainly exist in those nuclei where the GT model gives double peaks for GDR instead of only a single peak from experimental data, which can be seen in Fig. \ref{fig:fig4} for details. On the contrary, the neural networks give right classification of unimodal nuclei and bimodal nuclei. This can be seen in Table \ref{tab:accuracy}, where the accuracies of classification of unimodal nuclei and bimodal nuclei for neural network approach and GT model in different mass regions are shown. For nuclei with $A<170$ the neural network approach largely improves the accuracies of classification of unimodal nuclei and bimodal nuclei, leading to a great improvement in comparison with experimental data, being consistent with the results in Fig. \ref{fig:fig2}. This also explains the more considerable improvement by MTL approach for unimodal nuclei than that for bimodal nuclei observed in panel (a) of Fig. \ref{fig:fig2}. The wrong classification of unimodal nuclei and bimodal nuclei in GT model influences more peak energies than resonance widths, so the improvements by MTL are more apparent for energies of unimodal nuclei in Fig. \ref{fig:fig2}(a) than for widths in Fig. \ref{fig:fig2}(b).   In GT model, the GDR splits into two peaks for oscillations parallel to the axis of rotational symmetry and perpendicular to it in case of deformed nuclei. However, the wrong classification of unimodal nuclei and bimodal nuclei shows that the nuclear shape is not considered accurately in this model. Actually this is a general problem for phenomenological models. For example, in some empirical formulas in SLO method \cite{Capote2009}, the classification of unimodal nuclei and bimodal nuclei depends on quadrupole deformation parameter $\beta_2$, where the nucleus is considered as an unimodal nucleus when $\beta_2 < 0.01$ \cite{RIPL3}, and a bimodal nucleus otherwise.  So it is important for the description of GDR  if  the value of $\beta_2$ reflects the deformation of nucleus correctly, which is not always obvious. The general shape of deformed nuclei is an axially symmetric prolate or oblate ellipsoid with a deformation parameter $\beta_2$, which is defined by expanding the nuclear surface in spherical harmonics $R(\Omega) \approx R_0 (1+\beta_2 Y_{20} (\Omega))$, reflecting the difference between the nuclear radii along ($R_{\rVert}$) and perpendicular ($R_{\perp}$) to the symmetry axis \cite{Bohr1998, Harakeh2001}. Experimentally, the deformation parameter $\beta_2$ is extracted from the experimental reduced electric quadrupole transition probability B(E2) value through $\beta_2=(4\pi/3ZR_0^2)[B(E2)\uparrow/e^2]^{1/2}$, where $R_0 = 1.2 A^{1/3}$ fm \cite{Raman20011}. However, to indicate the presence of collective quadrupole effects in nuclei, this way to extract $\beta_2$ is sometimes less useful because it includes effects which vary with the size of nucleus (larger $\beta_2$ for light nuclei) \cite{Raman20011}. So the $\beta_2$ extracted this way does not always reflect the nuclear shape accurately, especially for light mass nuclei. On the other hand, by considering the dipole oscillation as a standing wave in a resonator, the oscillation frequency along the axes that is parallel to ($K=0$)  and perpendicular to ($K=1$) the symmetry axes is different, which is proportional to $1/R_{\rVert}$ and $1/R_{\perp}$, respectively, so that the splitting of these two frequencies $\Delta E \propto \beta_2$ \cite{Harakeh2001, Speth1981}, reflecting more accurately about nuclear shape. For example, for doubly magic nucleus $^{16}$O, it is considered as a spherical nucleus, which is reflected by the single peak of GDR observed from experiment \cite{Plujko2018}, although its experimental $\beta_2=0.364$ \cite{RIPL3}.

In Fig. \ref{fig:fig4}, we further list the misjudged nuclei by GT model, where the unimodal nuclei are considered as bimodal nuclei in GT model. The corresponding proton number of isotopes or neutron number of isotones is marked in Fig. \ref{fig:fig4}. It can be seen that these proton or neutron numbers are either (close to) magic numbers, such as 18 and 22 near 20, 28 and its neighbor 29, 50, and 79 near 82, or (close to) closed shells, such as 32, 40, 58 and 64. The nucleus with proton or neutron number close to magic number or big shell closure tends to have a spherical shape, and correspondingly the peak of GDR should not be split, which agrees with the experimental observation. So in our approach, we first establish a classification neural network (Net$_0$), with not only experimental $\beta_2$  but also  proton number $Z$, neutron number $N$ and mass number $A$ in the input layer.  As a consequence, the accuracy of Net$_0$ for classification of unimodal nuclei and bimodal nuclei can reach about $98\%$ with $98.3\%$ for learning set and $92.8\%$ for validation set. Thus, based on precise classification and combined with the MTL method, the description of GDR parameters has been improved obviously by networks, especially for unimodal nuclei.

\begin{figure}[htbp]\setlength{\abovecaptionskip}{0.0em}
\includegraphics[width = 0.99\linewidth]{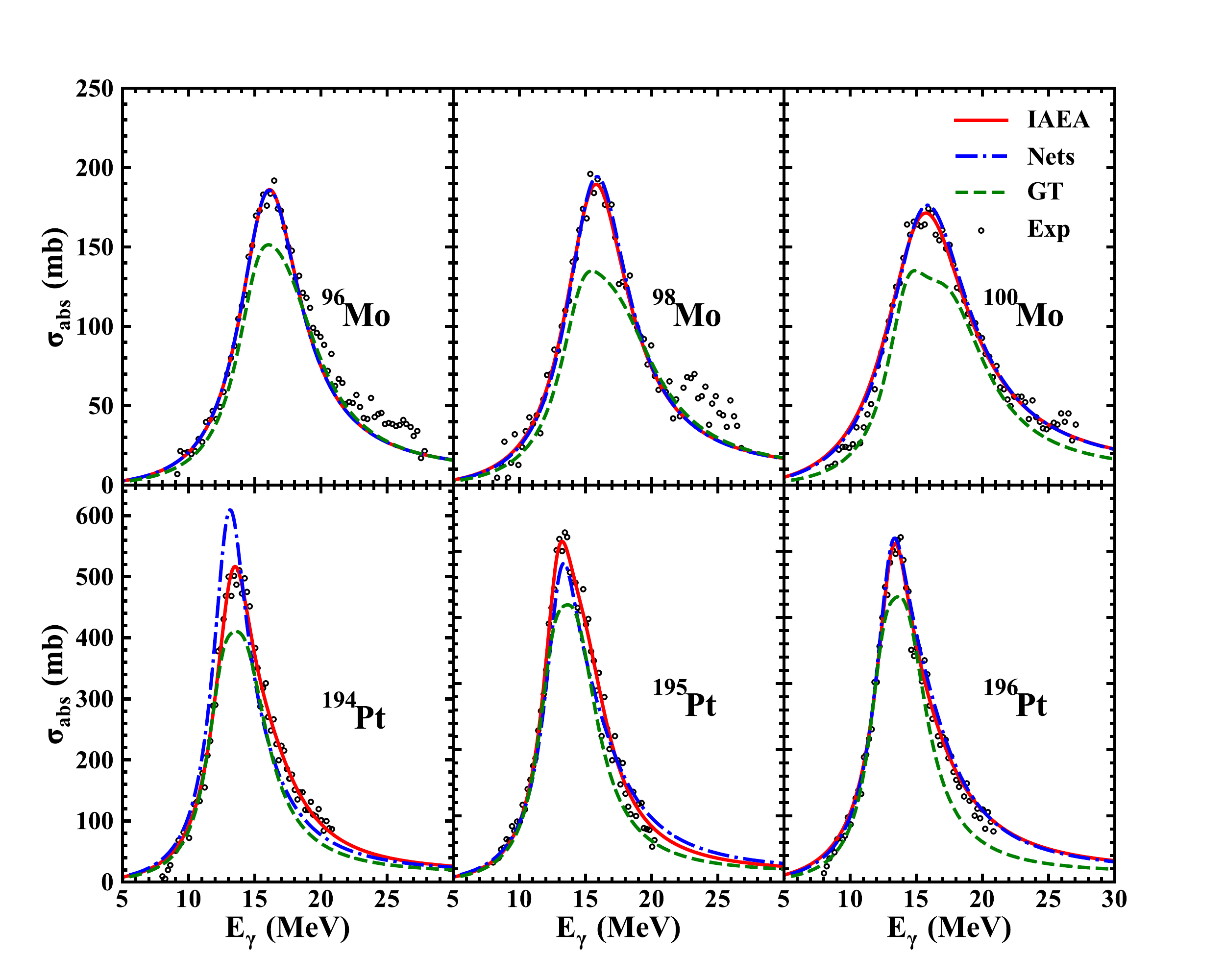}
\caption{
The photoabsorption cross-sections of $^{96,98,100}$Mo and $^{194,195,196}$Pt obtained by SLO method using GDR key parameters from MTL networks (dot-dashed blue), GT model \cite{RIPL3} (dashed green), and IAEA library \cite{Plujko2018} (solid red). The experimental data of photoabsorption cross-sections are also shown for comparison. These nuclei are included in the learning set.}\label{fig:fig5}
\end{figure}

With the GDR key parameters, one can obtain the photoabsorption cross-sections through SLO method, as introduced in Section 2. So in Fig. \ref{fig:fig5}, we plot the photoabsorption cross-sections obtained by SLO method using GDR key parameters from MTL networks, GT model, and IAEA library.  Since in the neural network approach, only energies and widths are studied, so here we still use $s_j$ from evaluated data \cite{Plujko2018}. The original experimental data are also shown for comparison. Here we choose the unimodal isotopes $^{96,98,100}$Mo and bimodal isotopes $^{194,195,196}$Pt in the learning set of our neural networks as examples. The results obtained by GDR parameters evaluated in IAEA library through SLO method have excellent agreement with the experimental data, especially in the GDR region. It further proofs that using IAEA evaluated data of GDR key parameters as neural networks' learning targets is good enough for describing photoabsorption cross sections. For the unimodal Mo isotopes, GT model considers them as bimodal nuclei inaccurately, leading to a big deviation from experimental data. The MTL networks give a good description of experimental data via right classification. For bimodal nuclei, although GT model gives a better performance than it does for unimodal nuclei, one still can see the clear improvement by MTL networks.

\begin{figure}[htbp]\setlength{\abovecaptionskip}{0.0em}
\includegraphics[width = 0.99\linewidth]{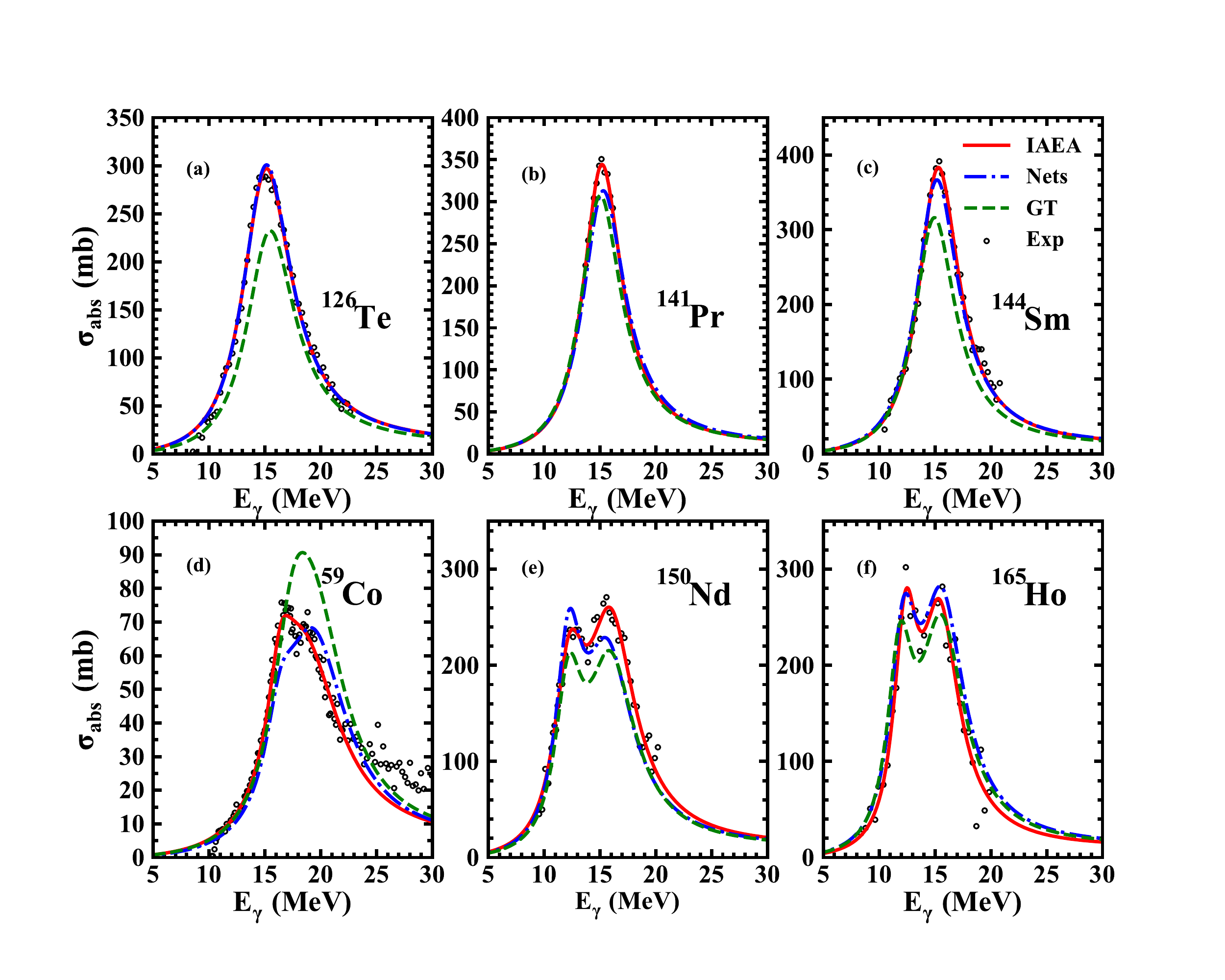}
\caption{Similar to Fig. \ref{fig:fig5} but for nuclei $^{126}$Te, $^{141}$Pr, $^{144}$Sm, $^{59}$Co, $^{150}$Nd and $^{165}$Ho which  are included in the validation set. }\label{fig:fig6}
\end{figure}

To check the extrapolation ability of MTL networks, we further study those nuclei in the validation set which are not taken into the learning progress. So in Fig. \ref{fig:fig6}, the photoabsorption cross-sections of 3 unimodal nuclei $^{126}$Te, $^{141}$Pr and $^{144}$Sm, and 3 bimodal nuclei $^{59}$Co, $^{150}$Nd and $^{165}$Ho in validation set are shown. It can be seen even for these nuclei in the validation set, the MTL networks still give a good prediction for the GDR key parameters. This result shows our ability for extrapolation, as well as the usefulness of our study for predicting photoabsorption cross-sections, especially for nuclei with few experimental data points, e.g. in panel (c), (e) and (f).

Based on the good performance in the validation set, we do further extrapolation of our approach to nuclei not included in IAEA library. Considering the reliability of extrapolation, we predict another 79 nuclei around the $\beta$-stability line, since in the learning set only stable nuclei are included.  These nuclei have no experimental data. The predicted peak energies and resonance widths of these nuclei are shown in Table \ref{tab:prediction} in appendix. There are 51 nuclei with unimodal distribution and 28 nuclei with bimodal distribution. This table provides a good reference for future data evaluations.

\section{Summary}

We introduced the MTL neural network approach to learn and predict GDR key parameters for the first time. The accuracy for the description of GDR key parameters is improved considerably compared to the theoretical GDR parameters calculated by GT model in RIPL-3. Especially, compared with GT model, for the GDR energies and widths of unimodal nuclei, the accuracies of MTL are about doubled,  for the bimodal nuclei, the accuracies of MTL are increased by about one third.  For the GDR energies of unimodal nuclei, the improvement by MTL is even more significant, which is due to the correct classification of unimodal and bimodal nuclei in the neural network approach. In GT model, the nuclear deformation is not considered accurately, resulting double peaks of GDR in some spherical nuclei with proton or neutron number close to magic number. The neural network approach overcomes this problem by introducing a network that classify the unimodal and bimodal nuclei, with an accuracy of about 98\%. Based on the good performance of both learning set and validation set of neural network approach, an extrapolation to 79 nuclei around the $\beta$-stability line without experimental data is made, which provides an important reference to future experiments and data evaluations.

As an improvement to the present work, to learn and predict experimental data of photoabsorption cross section directly using BNN approach is under progress.
In addition, the successful application of MTL approach in this work shows the feasibility of studying multi-output physical problems, so this approach can be generalized to other nuclear physics problems with multi-outputs.

\section*{Acknowledgement}
This research is supported by the National Natural Science Foundation of China No. 11875152, No. 12075104, No. 11875070, the Fundamental Research Funds for the Central Universities under Grant No. Lzujbky-2019-11, and the Open fund for Discipline Construction, Institute of Physical Science and Information Technology, Anhui University.

\begin{center}
\appendix
\section{Predictions}
\begin{longtable}{c@{\extracolsep{\fill}}ccccc}
\caption{Predictions of GDR key parameters including peak energies and resonance widths for nuclei without experimental data.} \label{tab:prediction}
\\
\multicolumn{6}{c}{} \\
\hline
Element & $\beta_2$ & $E_1$ & $\Gamma _1$ & $E_2$ & $\Gamma _2$  \\
\hline
\endfirsthead
\\
\multicolumn{6}{l}{TABLE \ref{tab:prediction} continued.} \\
\hline
Element & $\beta_2$ & $E_1$ & $\Gamma _1$ & $E_2$ & $\Gamma _2$  \\
\hline \endhead
\hline \endfoot
  \endlastfoot
        $^{20}$Ne & 0.364  & 23.016  & 5.738  &  &  \\
        $^{21}$Ne & 0.372  & 18.188  & 2.729  & 20.844  & 4.430  \\
        $^{22}$Ne & 0.384  & 16.938  & 2.857  & 21.328  & 5.102  \\
        $^{31}$P & 0.218  & 23.000  & 6.879  &  &  \\
        $^{33}$S & 0.209  & 22.766  & 7.027  &  &  \\
        $^{35}$Cl & 0.234  & 21.953  & 7.383  &  &  \\
        $^{37}$Cl & 0.011  & 21.438  & 5.836  &  &  \\
        $^{36}$Ar & 0.255  & 19.938  & 9.977  &  &  \\
        $^{38}$Ar & 0.000 & 21.120  &  5.805  &  &  \\
        $^{39}$K & 0.032  & 20.969  & 5.918  &  &  \\
        $^{41}$Ca & 0.021  & 20.453  & 5.855  &  &  \\
        $^{43}$Ca & 0.011  & 16.313  & 4.066  & 25.047  & 5.168  \\
        $^{45}$Sc & 0.043  & 19.734  & 5.859  &  &  \\
        $^{47}$Ti & 0.053  & 17.531  & 4.227  & 20.984  & 3.768  \\
        $^{49}$Ti & 0.053  & 16.984  & 4.031  & 20.344  & 5.500  \\
        $^{50}$Cr & 0.194  & 19.469  & 6.531  &  &  \\
        $^{53}$Mn & 0.021  & 16.750  & 3.918  & 19.891  & 6.555  \\
        $^{56}$Fe & 0.117  & 16.406  & 3.232  & 19.422  & 6.898  \\
        $^{57}$Fe & 0.162  & 16.406  & 3.230  & 19.422  & 6.898  \\
        $^{61}$Ni & 0.107  & 16.406  & 3.232  & 19.422  & 6.898  \\
        $^{62}$Ni & 0.107  & 16.406  & 3.232  & 19.422  & 6.898  \\
        $^{66}$Zn & 0.176  & 17.516  & 5.609  &  &  \\
        $^{67}$Zn & 0.176  & 17.406  & 5.598  &  &  \\
        $^{68}$Zn & 0.136  & 17.266  & 5.336  &  &  \\
        $^{69}$Ga & 0.177  & 17.297  & 5.523  &  &  \\
        $^{71}$Ga & 0.207  & 17.000  & 5.785  &  &  \\
        $^{83}$Kr & 0.129  & 16.641  & 5.016  &  &  \\
        $^{84}$Kr & 0.086  & 16.641  & 4.758  &  &  \\
        $^{85}$Rb & 0.064  & 16.672  & 4.625  &  &  \\
        $^{86}$Sr & 0.000 &  16.660 & 4.438  &  &  \\
        $^{87}$Sr & 0.043  & 16.656  & 4.527  &  &  \\
        $^{88}$Sr & 0.000 & 16.620 & 4.426  &  &  \\
        $^{97}$Mo & 0.172  & 15.938  & 5.879  &  &  \\
        $^{99}$Tc & 0.194  & 15.859  & 6.340  &  &  \\
        $^{101}$Ru & 0.195  & 15.836  & 6.387  &  &  \\
        $^{102}$Ru & 0.206  & 15.789  & 6.637  &  &  \\
        $^{104}$Pd & 0.173  & 15.906  & 5.977  &  &  \\
        $^{105}$Pd & 0.174  & 15.805  & 6.039  &  &  \\
        $^{106}$Pd & 0.185  & 15.719  & 6.207  &  &  \\
        $^{107}$Pd & 0.195  & 15.656  & 6.316  &  &  \\
        $^{110}$Cd & 0.152  & 15.711  & 5.762  &  &  \\
        $^{111}$Cd & 0.162  & 15.617  & 5.844  &  &  \\
        $^{113}$Cd & 0.185  & 15.484  & 5.977  &  &  \\
        $^{121}$Sb & 0.125  & 15.367  & 5.164  &  &  \\
        $^{123}$Te & 0.146  & 15.320  & 5.301  &  &  \\
        $^{125}$Te & 0.125  & 15.242  & 4.992  &  &  \\
        $^{129}$Xe & 0.162  & 15.148  & 5.664  &  &  \\
        $^{131}$Xe & 0.125  & 15.102  & 5.031  &  &  \\
        $^{132}$Xe & 0.125  & 15.055  & 5.078  &  &  \\
        $^{134}$Ba & 0.125  & 15.117  & 5.051  &  &  \\
        $^{135}$Ba & 0.125  & 15.070  & 5.094  &  &  \\
        $^{136}$Ba & 0.021  & 15.219  & 4.527  &  &  \\
        $^{137}$Ba & 0.053  & 15.102  & 4.363  &  &  \\
        $^{147}$Sm & 0.140  & 14.930  & 5.547  &  &  \\
        $^{149}$Sm & 0.183  & 14.539  & 6.379  &  &  \\
        $^{155}$Gd & 0.249  & 12.563  & 2.732  & 15.719  & 6.078  \\
        $^{157}$Gd & 0.271  & 12.266  & 2.586  & 15.805  & 5.930  \\
        $^{161}$Dy & 0.271  & 12.250  & 2.674  & 15.773  & 5.699  \\
        $^{163}$Dy & 0.283  & 12.211  & 2.936  & 15.680  & 4.828  \\
        $^{167}$Er & 0.297  & 12.211  & 3.053  & 15.617  & 4.313  \\
        $^{171}$Yb & 0.299  & 12.211  & 3.068  & 15.602  & 4.234  \\
        $^{173}$Yb & 0.300  & 12.211  & 3.076  & 15.594  & 4.199  \\
        $^{210}$Pb & 0.000 & 13.600 & 3.700  &  &  \\
        $^{212}$Bi & 0.011  & 13.594  & 3.752  &  &  \\
        $^{213}$Bi & 0.010  & 13.586  & 3.736  &  &  \\
        $^{214}$Bi & 0.010  & 13.578  & 3.727  &  &  \\
        $^{216}$Bi & 0.046  & 13.578  & 3.920  &  &  \\
        $^{218}$Po & 0.056  & 13.578  & 3.992  &  &  \\
        $^{221}$Rn & 0.110  & 13.063  & 2.617  & 13.883  & 5.684  \\
        $^{222}$Rn & 0.110  & 13.039  & 2.598  & 13.797  & 5.594  \\
        $^{223}$Fr & 0.132  & 12.719  & 2.451  & 13.047  & 4.488  \\
        $^{223}$Ra & 0.132  & 12.734  & 2.463  & 13.086  & 4.555  \\
        $^{224}$Ra & 0.143  & 12.555  & 2.275  & 12.695  & 3.609  \\
        $^{225}$Ra & 0.154  & 12.359  & 1.953  & 12.383  & 2.930  \\
        $^{226}$Ra & 0.164  & 12.109  & 1.504  & 12.211  & 3.057  \\
        $^{227}$Ac & 0.164  & 12.078  & 1.466  & 12.125  & 3.012  \\
        $^{228}$Th & 0.174  & 11.789  & 0.979  & 12.203  & 3.799  \\
        $^{229}$Th & 0.184  & 11.461  & 0.680  & 12.531  & 4.699  \\
        $^{230}$Th & 0.195  & 11.180  & 0.964  & 13.117  & 5.313  \\  \hline
\end{longtable}

\newpage

%\section*{References}
\end{center}
%\bibliographystyle{elsarticle-num}
%\bibliography{References}

\end{document}